\documentclass[pre, aps,twocolumn,nofootinbib,showpacs]{revtex4}
\usepackage{amsmath}
\usepackage{amsfonts}
\usepackage{amssymb}
\usepackage{graphics}
\usepackage{graphicx}
\usepackage{upgreek}
\usepackage{color}

\newcommand{\ii}{\text{i}}
\newcommand{\ud}{\text{d}}
\newcommand{\mus}{\mu\text{s}}
\newcommand{\mum}{\mu\text{m}}

\begin{document}

\title{Influence of positional correlations on the propagation of waves in a complex medium with polydisperse resonant scatterers}

\author{Valentin Leroy}
\email{valentin.leroy@univ-paris-diderot.fr}
\affiliation{Laboratoire MSC, Universit\'e Paris-Diderot, CNRS (UMR 7057), Paris, France}

\author{Anatoliy Strybulevych}
\affiliation{Department of Physics \& Astronomy, University of Manitoba, Winnipeg, Canada}

\author{Martin G. Scanlon}
\affiliation{Department of Food Science, University of Manitoba, Winnipeg, Canada}

\author{John H. Page}
\affiliation{Department of Physics \& Astronomy, University of Manitoba, Winnipeg, Canada}


\date{\today}

\begin{abstract}
We present experimental results on a model system for studying wave propagation in a complex medium exhibiting low frequency resonances. These experiments enable us to investigate a fundamental question that is relevant for many materials, such as metamaterials, where low-frequency scattering resonances strongly influence the effective medium properties.  This question concerns the effect of correlations in the positions of the scatterers on the coupling between their resonances, and hence on wave transport through the medium. To examine this question experimentally, we measure the effective medium wave number of acoustic waves in a sample made of bubbles embedded in an elastic matrix over a frequency range that includes the resonance frequency of the bubbles. The effective medium is highly dispersive, showing peaks in the attenuation and the phase velocity as functions of the frequency, which cannot be accurately described using the Independent Scattering Approximation (ISA). This discrepancy may be explained by the effects of the positional correlations of the scatterers, which we show to be dependent on the size of the scatterers. We propose a self-consistent approach for taking this ``polydisperse correlation'' into account and show that our model better describes the experimental results than the ISA. 
\end{abstract}

\maketitle

\section{Introduction}\label{section1}
Propagation of waves in complex media continues to be a very active subject of research. Of particular interest are complex media 
 {containing} 
 scatterers { with resonances at low frequencies}. Indeed, when the wavelength is large compared to the typical size of the inhomogeneities
, one can use an effective  {medium} approach, \emph{i.e.}, consider that the wave propagates as it would do in a homogeneous medium with an effective wavevector whose value is related to the composition and the structure of the 
 {material}. In most cases, the effective medium has properties that are close to the average of that of its components. But when there are resonant scatterers, their contribution to the wave field can be strong enough to significantly alter the propagation. In some cases, the effective media 
 show intriguing properties, such as a 
  negative refractive index, which are not encountered in nature; such materials are called \emph{metamaterials}~\cite{Pendry2000, Liu2000, Smith2004, Yang2008}.

One of the key issues for wave propagation in strongly scattering materials is the effect of coupling between the scatterers near resonance, an effect that makes the traditional independent scattering approximation (ISA) inadequate. Indeed, when the resonances are strong and/or the concentration of scatterers high, positional correlations and  {multiple scattering} loops may have a non-negligible contribution. To gain a better understanding of the underlying physics, experiments on a well-controlled system with strong resonances are needed.   {One such system can be found in acoustics, where } 
bubbly liquids are often regarded as model systems for the study of wave propagation with low frequency resonances. Indeed, air bubbles in a liquid have a particularly low resonance, know{n} as the Minnaert resonance, {with }
an angular frequency given by
$$
\omega_M = \frac{v_1\sqrt{3\rho_1/\rho_0}}{R},
$$
where $R$ is the bubble radius, $v_1$ is the velocity of longitudinal waves in 
 air, and $\rho_1$ and $\rho_0$ {are} the mass densities of the air and the liquid, respectively. As the ratios of the densities and 
  the velocities are small, it is easy to see that this resonance is at very low frequency:
$$
\frac{\omega_M R}{v_0} = \sqrt{3\rho_1/\rho_0}\frac{v_1}{v_0} \ll 1.
$$
Here, $v_0$ is the longitudinal sound velocity in the liquid.  This equation shows that, at resonance, the wavelength is large
compared to the bubble size.  This is important because, then, at the resonance frequency,  not only is the response of the bubble 
strong, but several bubbles (and potentially many) can be driven in phase, thus yielding a collective constructive contribution to the pressure field. This explains why a minute quantity of gas bubbles can dramatically change the effective acoustic properties of a liquid.

If the bubble is no longer in a liquid, but in an elastic medium with shear velocity $u_0$, its resonance frequency becomes~\cite{Alek1999,Zabolotskaya2005}:
\begin{eqnarray}
\omega_M= \frac{\sqrt{3v_1^2\rho_1/\rho_0+4u_0^2}}{R},
\label{omega0}
\end{eqnarray}
so the same condition of 
small radius {with respect to }
wavelength {persists }
as long as the elastic medium is a soft solid in which $u_0\ll v_0$. {As a result, }
elastic bubbly media are also good candidates as model systems for {investigating wave }
propagation in {the }presence of low frequency resonances. {In addition}
, they offer two practical {advantages }
compared to bubbly liquids: ($1$) bubbles do not move, which means that a precise knowledge of their positions can be obtained; ($2$) the frequency of the resonance can be shifted by tuning the shear velocity in the elastic matrix.
Bubbly soft media with negligible shear velocity have already been studied~\cite{Strybulevych2007,Leroy2008,Leroy2009}, and the ISA was found to 
 {reliably predict the experimental results, at least for concentrations of bubbles up to 1$\%$}. The present article focuses on a bubbly polydimethylsiloxane (PDMS) sample, in which {the }shear modulus is expected to significantly change the bubble's 
 {response to the incident ultrasonic field} and the 
propagation  {of waves in the medium}. Our aim is to  {see} 
if the ISA is still relevant in th{is }
case.
Beyond the fundamental aspect of the study, and the potential applications for design of metamaterials, it should be noted that a better understanding of the acoustic properties of bubbly elastic media is interesting \emph{per se}, since many practical applications, in the medical or the food science fields for instance,  {involve} 
the presence of bubbles in a soft elastic matrix.

The present paper is organized as follows.  {The next} 
section gives a brief overview of the ISA, with a focus on its application to 
acoustic propagation in bubbly media. Then we describe how our bubbly samples were made and characterized. Section~\ref{XP}  {reports} 
the experimental measurements  of velocity and attenuation in one typical sample, and compares the experimental results with the ISA. Two main limitations of the model are also discussed:  {the assumption of sample homogeneity on length scales larger than the scatterer size (e.g., gradients in the concentration of the scatterers) and the neglect of correlations in the positions of the scatterers.  Methods for overcoming these limitations are proposed.} 
\section{The Independent Scattering Approximation}\label{sec-foldy}
Based on a diagrammatic representation of scattering events~\cite{Sheng1995}, the ISA predicts the following effective wavenumber $k$ for a medium with $n$ scatterers per unit volume defined by their scattering function $f_\text{s}$:
\begin{equation}
k^2 = k_0^2 +4\pi n f_\text{s},\label{eqISA}
\end{equation}
where $k_0$ is the wavenumber in the pure medium (matrix). The ISA is actually equivalent to the model developed, with another approach, by Foldy in $1945$~\cite{Foldy1945}. In the following, we will use equally ``ISA'' or ``Foldy's model'' when refering to equation~(\ref{eqISA}).

In eq.~(\ref{eqISA}), {the effect of the scatterers on propagation in the pure medium is described by }
a  {term proportional to}  $n f_\text{s}$
, \emph{i.e.}, a simple addition of the individual scattering events, which are {assumed }
to be independent. Models have been proposed for incorporating either the correlations between the scatterings~\cite{Keller1964} or the loops, \emph{i.e.}, the events in which the same scatterer  {or spot inside the sample} is visited  {more than once} 
by the wave~\cite{Hen99, Haney2003}. We discuss only the ISA  {in this section}
, since it is the simplest model and it {gives }
reasonable predictions  {of wave transport} in many cases. We will 
introduce {later}, in section~\ref{sec-correlations} and the appendix, extensions to the ISA for dealing with the positional correlations of the scatterers.

Equation~(\ref{eqISA}) considers a monodisperse assembly of scatterers. In practical applications, one is often dealing with polydisperse media, in which case the ISA predicts that
\begin{equation}
k^2 =k_0^2 + \int 4\pi n(R) \text{d}R f_s(R),\label{Eqfoldy}
\end{equation}
where $n(R)\text{d}R$ is the number of bubbles per unit volume whose radius is between $R$ and $R+\text{d}R$, and $f_s(R)$ is the scattering function for a scatterer of radius $R$.

Let us apply the ISA to a bubbly medium. When the scatterer is a bubble, its isotropic scattering function is given by
\begin{equation}
f_s(\omega,R) = \frac{R}{(\omega_M/\omega)^2 -1 + \ii \delta},\label{eqf}
\end{equation}
where $\delta$ is the damping constant due to thermal, viscous and radiative losses~\cite{pro77} and $\omega_M$ the resonance angular frequency given by Eq.~(\ref{omega0}), which can be written
\begin{eqnarray}\label{MinnaShear}
\omega_M^2 = \frac{3\kappa P_1 + 4\mu^\prime}{\rho_0 R^2},
\end{eqnarray}
with $\kappa$ the polytropic index for the transformations undergone by the gas~\cite{pro77} ($\kappa=1.4$ 
for air), $P_1$ the pressure of the gas in the bubbles, $\rho_0$ the mass density of the matrix, and $\mu^\prime$ the real part of the shear modulus of the matrix.
Surface tension effects are negligible for the bubble size considered in our experiments. Mode conversion is not considered in this model since it has been shown to be negligible for this system on account of the large difference between the shear modulus of the matrix and the longitudinal one ($u_0\ll v_0$)~\cite{Liang2007}. Hence the only significant contribution of the nonzero shear modulus is its effect on the resonance frequency of the bubble, which is increased by the stiffness of the matrix. Another approximation we make is to reduce the scattering function to its isotropic component, neglecting the dipolar and higher contributions. This approximation is generally good when the wavelength is large compared to the size of the scatterers, and it will be shown to be accurate for describing our experiments.

\begin{figure}[hbt]
	\centering
	\includegraphics[width = \linewidth]{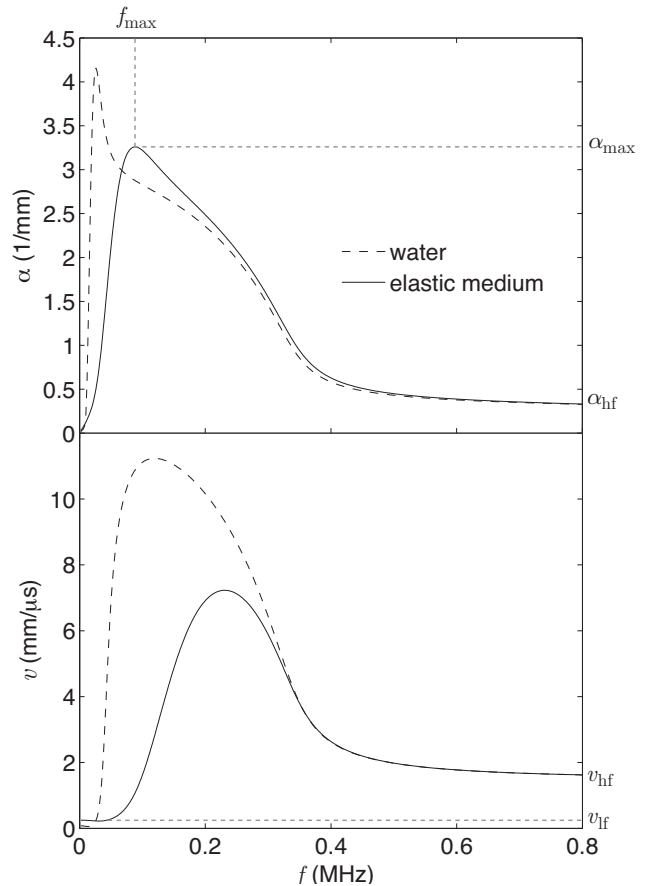}	
   	\caption{\textit{(Color online) Prediction{s } of the ISA for the 
    attenuation $\alpha$ (top) and phase velocity $v$ (bottom) as functions of 
    frequency for bubbles in water (dashed lines) and in an elastic matrix (solid lines) with $\mu^\prime = 0.7\,$MPa, $\mu^{\prime\prime}=0.4\,$MPa; {the bubble }
    concentration is $2\,\%$ and the mean radius is taken {to be }
    $150\,\mum$ with a $25\,\%$ polydispersity. The characteristic features ($f_\text{max}$, $\alpha_\text{max}$, $\alpha_\text{hf}$, $v_\text{lf}$, and $v_\text{hf}$) are indicated.}}
	\label{fig_simple}
\end{figure}

In figure~\ref{fig_simple}, we give two examples of the effective attenuation and velocity as functions of the frequency, as calculated by Foldy's model. For the sake of generality, we consider log-normal distributions of bubble size:
\begin{eqnarray}
n(R)=\frac{n_0}{\sqrt{2\pi}\epsilon R}\text{exp}\left(-\frac{(\text{ln}R/R_0)^2}{2\epsilon^2}\right),
\end{eqnarray}
where $R_0$ is the median radius, $\epsilon$ the width of the distribution, and $n_0$ the bubble concentration. Note that for distributions that are not too broad ($\epsilon <0.2$), log-normal distributions are close to normal distributions: then $R_0$ and $\epsilon$ can be viewed as the mean radius and 
 {coefficient of variation} (\emph{i.e.}, standard deviation divided by the mean), respectively.   {The latter quantity, which we will refer to as the polydispersity factor, measures the range of bubble sizes in the distribution.}  
 The bubble concentration $n_0$ is related to the gas volume fraction $\Phi$ by
\begin{eqnarray}
n_0 = \frac{3\Phi}{4\pi R_0^3 \text{exp}(9\epsilon^2/2)}.
\end{eqnarray}
The bubble size distribution in 
{Fig.}~\ref{fig_simple} is described by $R_0=150\,\mum$, $\epsilon=25\,\%$ and $\Phi=2\,\%$. Two host media are considered: water (dashed lines), and an elastic matrix with $\mu^\prime = 0.7\,$MPa, $\mu^{\prime\prime}=0.4\,$MPa (solid lines). For both media{, } very dispersive behavio
 r is predicted: the attenuation and velocity show peaks, which occur at higher frequencies in the elastic medium.

At low frequencies ($\omega<\omega_M$), sound propagation in the bubbly medium is weakly attenuated, and its very low phase velocity is accurately predicted by Wood's relation~\cite{Wood}, modified to account for the elastic effects. For $5{\times}
10^{-4}<\Phi<10^{-1}${, this relation can be approximated} 
(within $5\,\%$) {as }
\begin{eqnarray}
v_\text{lf} \simeq \sqrt{\frac{\kappa P_1 + 4\mu^\prime/3}{\rho_0 \Phi}},
\label{eq-vlf}
\end{eqnarray}
where $P_1$ is the pressure of the gas in the bubbles and $\rho_0$ the density of the matrix. Eq.~(\ref{eq-vlf}) gives, in our example, $v_\text{lf} = 122$ and $248\,$m/s in water and the elastic matrix, respectively.

Near the Minnaert frequencies of the bubbles, the acoustic waves are highly attenuated by the bubbly medium: the attenuation reaches a peak
 $\alpha_\text{max}$ at frequency $f_\text{max}$, corresponding to the resonance frequency for a monodisperse distribution. In water, the peak is very sharp and associated with the bubble resonances ($f_\text{max}=22\,$kHz as expected by Eq.~(\ref{MinnaShear}) for a $150$-$\mum$-radius bubble). Eqs.~(\ref{eqISA}) and (\ref{eqf}) give the physical origin of this sharp peak: at resonance, the scattering function $f_s$ is large and imaginary, which leads to a large imaginary part of the effective wave number $k$. This sharp resonant effect is followed by a broader regime (roughly $60$-$400\,$kHz in our example) in which the attenuation is still large. This is explained by the negative response of the bubbles. Indeed, when driven at a frequency higher than its resonance frequency, a bubble has a scattering function which reduces approximately to $f_s = -R$ when $\delta\ll 1$(see Eq.~(\ref{eqf})). Even though the bubble is not resonating, its contribution is far from negligible. Indeed, one can rewrite Eq.~(\ref{eqISA}) as
 \begin{equation}
 k^2=k_0^2 \left( 1 - \frac{R \lambda^2}{\pi d^3} \right),
 \end{equation}
 where $d$ is the typical distance between bubbles. As long as the wavelength is large compared to this distance ($\lambda/d\gg 1$), the square of the effective wavenumber is negative: waves are evanescent. In our example of $2\,\%$ bubbly water with $150\,\mum$-radius bubbles, the typical distance between bubbles is $0.9\,$mm, which means that $R \lambda^2/(\pi d^3)>1$ for $f<380\,$kHz.
 
In our example of an elastic medium, the high viscosity ($\mu^{\prime\prime}$ of $0.4\,$MPa corresponds, at $0.1\,$MHz, to a viscosity $640$ times larger than that of water) tends to damp the oscillations of the bubbles at resonance, so the attenuation peak is lower and less sharp (see figure~\ref{fig_simple}). As a paradoxical consequence, the more viscous the medium in which bubbles are embeded, the lower the maximum of attenuation of sound. Note that the sharp peak of attenuation also dissapears when the bubble size distribution is too broad. Interestingly, even if the resonant effects are too weak to give rise to significant resonant attenuation, they still lead to an evanescent regime. In this case, the position of the attenuation peak does not correspond to the resonance of the average size bubbles, but rather to a frequency that is just above the resonance frequency of the small bubbles of the distribution, as it corresponds to all the bubbles being driven at a frequency higher than their resonance frequencies.
  For instance, in our example $f_\text{max}=90\,$kHz for the bubbly elastic medium, which corresponds to a radius of $100\,\mum$ according to Eq.~(\ref{MinnaShear}). 
  
  When the resonances are smoothed out, an approximate formula can be established for the maximum of the attenuation:
 \begin{eqnarray}
 \alpha_\text{max} = \frac{2\sqrt{3\Phi}}{R_0 \text{exp}(2\epsilon^2)},
 \label{eq-amax}
 \end{eqnarray}
which has the benefit of depending only on the bubble concentration and size distribution ({not on }
either 
the damping constant of the bubbles 
or 
the shear modulus of the matrix). 

At higher frequencies, the evanescent regime disappears: the attenuation and velocity reach constant values, which can be approximated by:
\begin{eqnarray}
\alpha_\text{hf} &=& \frac{3\Phi}{R_0 \text{exp}(5\epsilon^2/2)},\label{eq-ahf} \\
v_\text{hf} &=& v_0.\label{eq-vhf}
\end{eqnarray}

Interestingly, if Foldy's model applies, one can determine the $5$ parameters $\Phi$, $R_0$, $\epsilon$, $\mu^\prime$, and $v_0$ from the measurement of the $5$ quantities $v_\text{lf}$, $f_\text{max}$, $\alpha_\text{max}$, $\alpha_\text{hf}$, and $v_\text{hf}$ in a bubbly medium.

\section{Materials and methods}
\subsection{Sample preparation}
Samples were prepared by mixing $100\,$g of RTV $615$ (GE Silicones, $90\,$g monomers and $10\,$g hardener) in a $600\,$mL beaker with a {Sorvall Omni Mixer (Ivan Sorvall, Inc., Norwalk, CT; nominal operating speed 16,000 rpm)} 
for $1\,$ minute to incorporate bubbles. Then, after waiting for an interval of $10$-$20\,$ minutes for the biggest bubbles to rise and disappear, the bubbly liquid was poured into a cell (thickness $2.6\,$mm, diameter $11\,$cm), which was rotated at $5\,$rpm to prevent creaming (bubbles rising to the top of the container). At room temperature, a solid sample was obtained in $1$ day.
\begin{figure}[hbt]
	\centering
	\includegraphics[width = .8\linewidth]{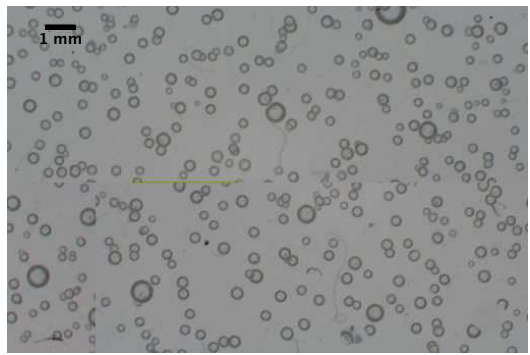}	
   	\caption{\textit{(Color online) Example of an image taken for the {bubble }size analysis (optical characterization).}}
	\label{fig_photo}
\end{figure}

\subsection{Sample characterization}\label{sec-charac}
\begin{figure}[hbt]
	\centering
	\includegraphics[width = \linewidth]{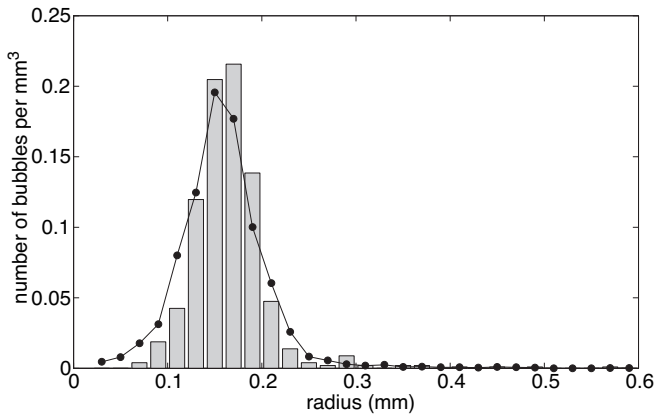}	
	\caption{\textit{(Color online) Histogram of the {bubble radius }
distribution for the sample in figure~\ref{fig_photo}
, obtained from the image analysis. The superposed curve {(circles plus connecting lines)} corresponds to the result of the x-ray tomography analysis.}}
	\label{fig_histo}
   \end{figure}

Testing of Foldy's model {requires }
that some of the parameters of the sample are known: the bubble size distribution ($n(R)$), the matrix rheology ($\mu$) and the velocity of sound in the matrix ($v_0$).
\paragraph{Optical characterization}was made by taking pictures at different positions in the sample. The depth of field was larger than the thickness of the sample, which means that the total volume of the image was known ($16\times11\times2.6\,$mm$^3$). A typical image is shown in figure~\ref{fig_photo}; the contrast was good enough for an automatic size analysis to be performed (ImageJ). The radius measurement was possible with a one pixel accuracy, \textit{i.e.}, $5\,\mum$. A total of $839$ bubbles was analyzed. To avoid biased measurements of big bubbles, overlaid bubbles were excluded from the size analysis. However, the total number of bubbles was determined ($1098$) for a correct estimation of the bubble volume fraction, assuming the same size distribution for the overlaid bubbles. The radius distribution was centred around $165\pm5\,\mum$ with a polydispersity of $21\pm1\,\%$ and a volume fraction of $2\pm0.2\,\%$ (see histogram in figure~\ref{fig_histo}).

\paragraph{Rheological characterization} was necessary to measure the shear modulus over the frequency range used for the ultrasonic experiments ($30$-$800\,$kHz). We used a shear wave reflection technique (see \cite{Longin1998} or \cite{Leroy2010} for details) to measure the complex shear modulus $\mu=\mu^\prime +\ii\mu^{\prime\prime}$ from $300$ to $500\,$kHz. For lower frequencies (around $30\,$kHz) an estimation of $\mu$ was possible by measuring the acoustic response of a single bubble (see~\cite{Strybulevych2009} for a description of the technique). Over the $30$-$500\,$kHz frequency range, our measurements are consistent with the following behaviour of the real ($\mu^\prime$) and imaginary ($\mu^{\prime\prime}$) parts of the shear modulus
:
\begin{subequations}\label{eqMU}
\begin{eqnarray}
\mu^\prime &=& 0.6 + 0.7\times f, \\
\mu^{\prime\prime} &=& 0.2 + 1.8\times f,
\end{eqnarray}
\end{subequations}
where the moduli are in MPa and the frequency $f$ in MHz. Equations~(\ref{eqMU}) only give the order of magnitude of the shear modulus for a given RTV$615$ sample. Indeed, the precise protocol used for preparation, such as the temperature and time of the hardening process for instance, may have a non-negligible impact on the modulus~\cite{Schneider2008}. As a consequence, deviations from the predictions of Eqs.~(\ref{eqMU}) can be observed for a given sample.

The stationary limit ($f=0\,$ MHz) of our estimate of $\mu^\prime$ is in 
good agreement with the value reported in the literature~\cite{Schneider2008}.

\paragraph{Ultrasonic characterization} was also performed by standard techniques~\cite{Leroy2008} to measure the velocity and attenuation of sound in the \textit{pure} PDMS. The longitudinal velocity was found to be constant over the frequency range investigated ($30\,$kHz - $5\,$MHz): $v_0=1.02\,$mm/$\mus$, in accordance with previously reported values~\cite{Delides1979}. For the attenuation{,} the following law was found: $\alpha_0=0.023f^{1.54}\,\text{mm}^{-1}$ (with frequency $f$ in MHz).

\paragraph{X-ray characterization.} To investigate the spatial distribution of the bubbles in one of the samples, x-ray tomography images were acquired, using an Xradia MicroXCT $3$D x-ray transmission microscope ($18\,\mum$/px). After the ultrasonic measurement{s}, the sample was cut in pieces small enough to fit in the apparatus ($7.5\times 10\times 2.6\,\text{mm}^3$), and $21$ pieces were imaged, for a total of $3700$ bubbles. A {software program was written to enable the reconstruction of }
the $3$D structure of each piece: for every bubble i, its radius $R_\text{i}$ and position $[x_\text{i}, y_\text{i}, z_\text{i}]$ were determined with 
$1$ pixel accuracy. The size distribution was determined 
and successfully compared to the results of the optical characterization: we found  a mean radius of  $160\pm18\,\mum$ with a polydispersity of $25\pm1\,\%$ and a volume fraction of $2.2\pm0.3\,\%$ (see figure~\ref{fig_histo} and table~\ref{Tab1}). Moreover, the $3$D data gave insight into the homogeneity of the sample, as well as on the correlations in the bubble positions, as described in sections~\ref{sec-compaLayer} and \ref{sec-correlations}.
\begin{table}[htdp]
\caption{\textit{Parameters of the bubble size distribution, extracted from the optical and x-ray measurements.}}
\begin{center}
\begin{tabular}{|l|c|c|c|}
\hline
  & {volume fraction} & {average radius}  & {polydispersity}  \\
  &  $\Phi$ ($\%$) &  $\langle R \rangle$ ($\mum$)  &   $\frac{\sqrt{\langle R^2 \rangle - \langle R \rangle^2}}{\langle R \rangle}$  ($\%$)\\
\hline
 optics & $2.0\pm0.2$ & $165\pm5$ & $21\pm1$ \\
x-ray & $2.2\pm0.3$     &  $160\pm18$ & $25\pm1$\\
\hline
\end{tabular}
\end{center}
\label{Tab1}
\end{table}

\subsection{Ultrasonic measurements}
The acoustic properties of the samples were measured with the following
set-up. In a large tank
($60\times60\times120\,\text{cm}^3$) filled with reverse osmosis
water,
a piezoelectric transducer generated a pulse that propagated
through water, traversed the sample and was detected by a hydrophone.
Because the attenuation in the sample was large, and because the divergence of the beam was not negligible (especially at low frequencies), the use of a screen was {essential }
to reduce spurious signals. The aperture of the screen ($6\,\text{cm}$) was larger than the wavelength of the pulse ($3.75\,\text{cm}$ at $40\,\text{kHz}$) to limit diffraction effects, but smaller than the diameter of the sample ($11\,$cm).

Gaussian pulses, with central frequencies ranging from $40$ to
$600\,\text{kHz}$, were generated by an Arbitrary Wave Generator,
and three different transducers, having central frequencies of $100$, $250$
and $500\,\text{kHz}$, were used to cover the range of
frequencies. The pulses were recorded, with a hydrophone, in two
different cases: when the sample was mounted on the screen
($s_1(t)$), and when the sample was absent ($s_2(t)$). The signals
were averaged over $100$ acquisitions when the attenuation was low
(for reference measurements, for example), and up to $5000$
acquisitions for highly attenuated signals. Signals $s_1(t)$ and
$s_2(t)$ were then truncated to eliminate spurious echoes, and Fourier transformed into $S_1(\omega)$ and
$S_2(\omega)$ respectively{.  Then}, 
$T(\omega)$, the ratio of the
transmission with and without the sample in the path of the acoustic
beam at a given angular frequency, was calculated.

For an incoming plane monochromatic wave $\mathrm{exp}(\ii
k x - \ii\omega t)$, $T(\omega)$ is given by
\begin{eqnarray}
T(\omega) = \frac{4ZZ_\text{w}}{(Z_\text{w}+Z)^2}\times \frac{\mathrm{e}^{\ii
(k-k_\text{w})d}}{1-\left(\frac{Z-Z_\text{w}}{Z+Z_\text{w}}\mathrm{e}^{\ii
kd}\right)^2 },
\label{eqT}
\end{eqnarray}
where $k$ and $k_\text{w}$ are the wave numbers in the sample and in the water respectively, $Z$ and $Z_\text{w}$ the acoustic impedances, and $d$ the thickness of the sample. Because the impedance $Z$ depends on $k$ ($Z=\rho\omega/k$, with $\rho$ the mass density of the sample), equation~(\ref{eqT}) is not directly invertible to extract $k$: an iteration method was used~\cite{Leroy2008}. From $k$, the phase velocity $v=\omega / \text{Re}(k)$ and the attenuation $\alpha = 2\text{Im}(k)$ can be calculated. Error-bars on the measurements were evaluated by taking into account the following sources of uncertainty: thickness of the sample, noise in the Fourier transforms, truncation of the time pulse.

An important issue in measuring the 
acoustic properties of a medium is the statistical relevance of {results for }a single sample. 
{T}heories predict the average properties, \emph{i.e.}, what one can measure when averaging over different samples having the same general parameters. To obtain insight on this issue, we measured the transmitted signal not only in the centre of the acoustic beam, but also on a $7\times 7$ grid with a $5\,$mm step. From this scanning, a ``typically averaged signal'' could be obtained (with correction for the geometrical differences), and this was compared to the centre pulse.
\section{Experimental results and discussion}\label{XP}
\begin{figure}[ht]
\begin{center}
\includegraphics[width = \linewidth]{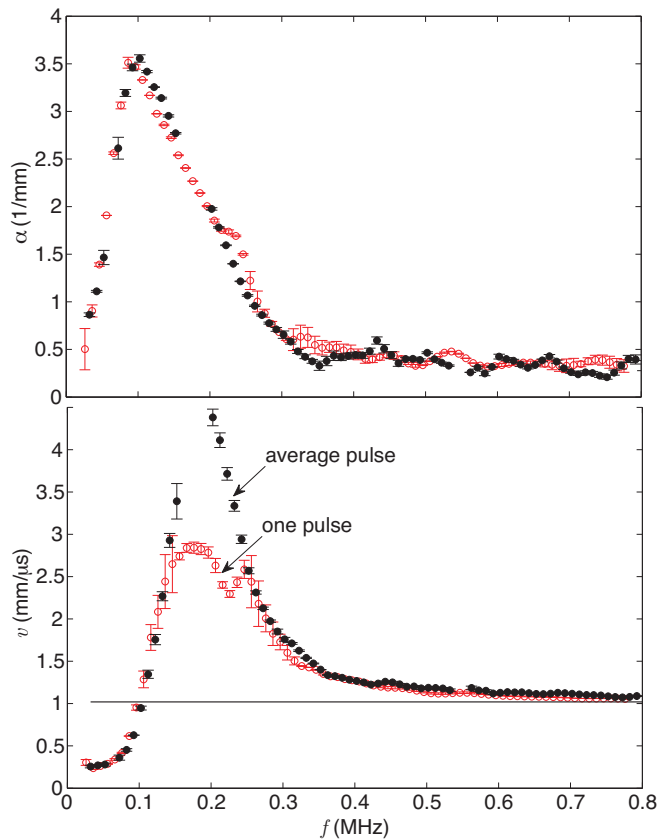}
\caption{\textit{(Color online) Attenuation $\alpha$ and phase velocity $v$ in the bubbly PDMS sample, measured from a single pulse (open circles) and from an average pulse (solid circles). The velocity in the pure PDMS is shown by the solid line. The attenuation in pure PDMS is, on average, $2$ orders of magnitude less than in the bubbly sample (too small to be visible on the graph).}\label{fig-PVAt}}
\end{center}
\end{figure}

Several bubbly PDMS samples were tested, with different void fractions ($0.5$-$3.5\,\%$) and polydispersities ($20$-$50\,\%$), but with the same average radius (around $150\,\mum$). In this article, we focus on representative results obtained with one selected sample, which was characterized by x-ray tomography. Figure~\ref{fig-PVAt} shows the measured attenuation and phase velocity in the sample as functions of the frequency. The general aspect of the curves follows what one would expect for a bubbly medium (see figure~\ref{fig_simple}): both the attenuation and the velocity exhibit peaks in frequency with very large values at their maxima ($3.5\,\text{mm}^{-1}$ and $4.5\,\text{mm/}\mus$, respectively), the velocity is low ($0.26\,\text{mm/}\mus$) at low frequencies, and close to the value in the pure PDMS  at higher frequencies.  Note that two cases are considered: when only one pulse is analysed (open circles) and when the average is made over the $49$ positions of the hydrophone (solid circles). Except for the peak in velocity, both analyses give similar results. It indicates that substantial information can be obtained from a single pulse measurement, which is interesting for practical purposes. Note that measurements are difficult when the velocity is high because small time shifts have to be extracted from the phase shift which is dominated by the complex impedance rather than the transit time.

The five typical quantities introduced in section~\ref{sec-foldy} can be measured and compared to equations~(\ref{MinnaShear}), and (\ref{eq-vlf})-(\ref{eq-vhf}), calculated with the parameters found in section~\ref{sec-charac}. 
As shown in table~\ref{Tab2}, {the }agreement is rather satisfying: $f_\text{max}$ and $\alpha_{hf}$ are within $10\%$ and the other quantities within $20\%$.  Note that for $v_\text{lf}$ and $v_\text{hf}$ the measured values may not correspond exactly to the asymptotic values because of the limited range of frequencies in the measurements.

\begin{table}[htdp]
\caption{\textit{Five characteristic quantities measured experimentally (first line), and predicted by the model (second line).}}
\begin{center}
\begin{tabular}{|l|c|c|c|c|c|}
\hline
  & $f_\text{max}$  & $\alpha_\text{max}$  & $\alpha_\text{hf}$ & $v_\text{lf}$& $v_\text{hf}$\\
 & (kHz) &  (mm$^{-1}$) &  (mm$^{-1}$) & (m/s)&  (m/s)\\
\hline
exp.  & $100$ & $3.5$ & $0.40$ & $260$ & $1180$\\
mod. & $90$\footnote{Calculated with Eq.~(\ref{MinnaShear}) by taking the radius of the small bubbles of the distribution (\emph{i.e.}, $100\,\mum$, see figure~\ref{fig_histo}).  This is appropriate since for polydisperse distributions of sizes, the peak in attenuation corresponds to the resonances of the smaller bubbles, as noted in section~\ref{sec-foldy}.}   & $2.8$ &  $0.35$       &  $224$       &  $1020$\\
\hline
\end{tabular}
\end{center}
\label{Tab2}
\end{table}%

\subsection{Comparison with ISA}

\begin{figure}[hbt]
\begin{center}
\includegraphics[width = \linewidth]{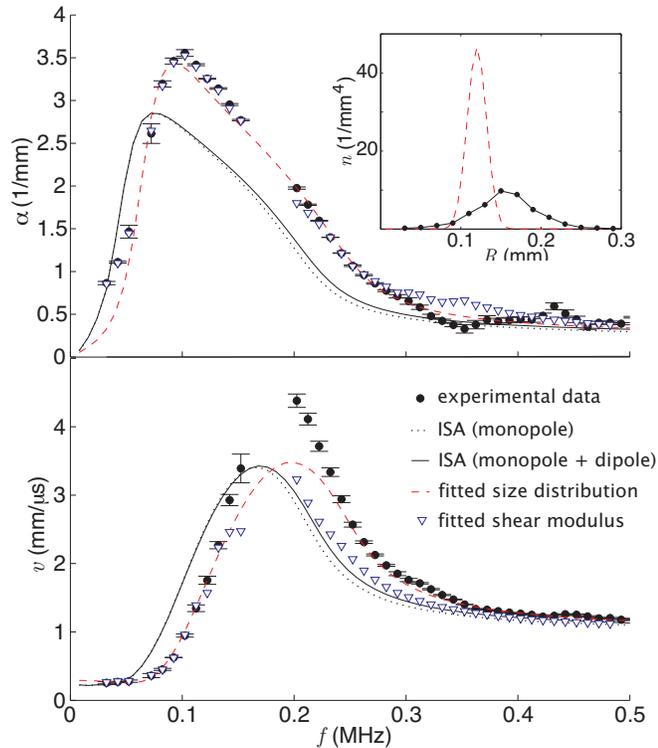}
\caption{\textit{(Color online) Close-up {of the attenuation and velocity peaks. }
Solid circles correspond to experimental data (same as in {Fig.}
~\ref{fig-PVAt}). Comparison with Foldy's model is shown for monopolar scattering of the bubbles (dotted lines) and monopolar + dipolar scattering (solid lines). {Note that there are no adjustable parameters in this comparison of the theory with experiment. }Results of 
fitting {the model to the data by allowing either} 
the size distribution (dashed lines) or the shear modulus (triangles) {to vary} are also shown. In the inset, the size distribution determined by the x-ray experiment (solid line) is compared to the 
one {inferred from the fit to the ultrasonic data} (dashed lines).}\label{fig-fitPVAt}}
\end{center}
\end{figure}

Figure~\ref{fig-fitPVAt} offers a further step in the comparison process. Predictions of the {ISA} model are plotted as dotted lines. Parameters for the model were taken as measured in section~\ref{sec-charac}, the histogram measured in the x-ray experiment being chosen for $n(R)$. In the low and high limits of the frequency range, the comparison is good. However, the attenuation and velocity peaks are not {correctly} 
predicted: their {frequencies} 
and amplitudes are both smaller than the experimental data. Note that this discrepancy was observed for all the samples we investigated.
The comparison is slightly better when the dipole terms in the bubbles' responses are considered (solid lines) but the discrepancy in the peaks' positions and magnitude is still large. Higher order 
scattering has negligible contributions.

To check whether the discrepancy could be explained by the uncertainty in the measured size distribution, we determined the $n(R)$ function that was needed to achieve a better fit. As shown in the inset of 
{Fig.}~\ref{fig-fitPVAt}, this leads to a totally unrealistic distribution, with smaller bubbles (mean radius $120\,\mum$), a narrower distribution ($10\,\%$ polydispersity), and a smaller void fraction ($1.1\,\%$). One can understand why such a size distribution is obtained by looking at equation~(\ref{eqISA}). If the peaks are at too low frequencies and with too low amplitudes, it means that the $nf_\text{s}$ term is not high enough compared to $k_0^2$, \emph{i.e.}, that the bubble contribution is too weak. A way of fixing this is to consider smaller bubbles, in order to shift the resonance to higher frequencies, and to increase the number of bubbles per unit volume to magnify their contribution. This explains the difference between the fitted and the actual size distribution in figure~\ref{fig-fitPVAt} inset. An important practical application is that if one uses Foldy's model for fitting the experimental data, an incorrect bubble size distribution will be obtained. Note that the same phenomenon was observed for bubbles in bread dough, with the same trend of Foldy's model predicting bubbles smaller than those determined from x-ray measurements~\cite{BiF2}.

\begin{figure}[ht]
\begin{center}
\includegraphics[width =\linewidth]{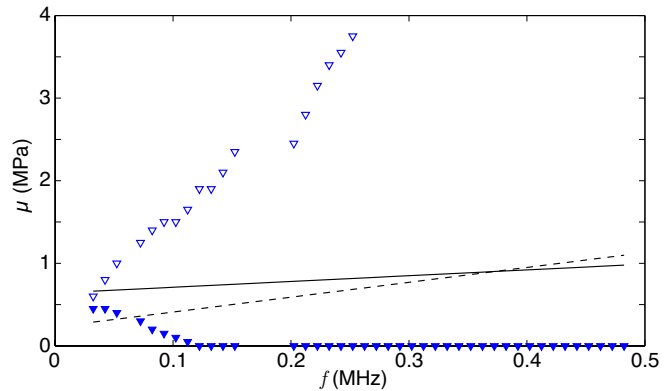}
\caption{\textit{(Color online) Shear moduli as functions of frequency. The lines correspond to the results for experimental $\mu^\prime$ (solid line) and $\mu^{\prime\prime}$ (dashed line), as given in equations~(\ref{eqMU}). Open triangles show the fitted values for $\mu^\prime$, solid ones for $\mu^{\prime\prime}$.}\label{fig-mufit}}
\end{center}
\end{figure}

Another possible source of error was the shear modulus, whose measurement was not very precise. For each frequency, we determined the complex shear modulus that gave the best fit 
 to the experimental data. As shown by the triangles in {Fig.}
 ~\ref{fig-fitPVAt}, a good fit was possible for the attenuation. For the velocity, the agreement was good for lower frequencies, but poor at higher frequencies. Moreover, as shown in figure~\ref{fig-mufit}, the fitting leads to unrealistic values for $\mu^\prime$ and $\mu^{\prime\prime}$. Again, one can qualitatively understand these values: the real part of the fitted shear modulus is high in order to shift the peaks to higher frequencies, and the imaginary part is low (even zero) so that the oscillations of the bubbles are not damped by viscosity, and the scattering function $f_\text{s}$ is large.

These results suggest that Foldy's model is inadequate for {describing wave propagation in }the PDMS sample. In the following we examine two possible explanations{, and propose a model to account for these discrepancies}.

\subsection{Homogeneity of the sample}\label{sec-compaLayer}
\begin{figure}[ht]
\begin{center}
\includegraphics[width = \linewidth]{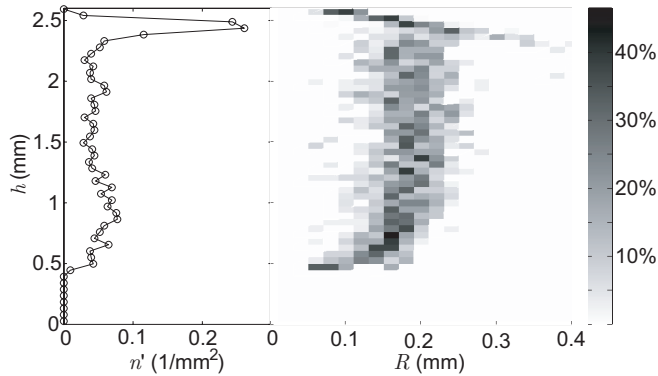}
\caption{\textit{(Color online) The layered structure of the sample is revealed by the x-ray tomography. Along the thickness $h$ of the sample, we report the number of bubbles per unit area $n^\prime$ (left) and the radii distribution histogram (right) for each layer.}\label{fig-homo}}
\end{center}
\end{figure}

An important hypothesis made in all the effective medium models consists in assuming that the sample is homogeneous, \textit{i.e.}, any part of the sample can be considered as equivalent to the others. The x-ray inspection of the sample reveals a homogeneous structure in the transverse directions, but a non-homogeneous one along the thickness. Probably because of the upward buoyancy force exerted on the bubbles during the filling of the cell, the concentration of bubbles is varying, as shown in {Fig.}
~\ref{fig-homo}: the bottom part of the sample is free of bubbles, whereas the top part {has a higher concentration. }
The bubble size distribution is also slightly different from one layer to the other.

\begin{figure}[ht]
\begin{center}
\includegraphics[width = \linewidth]{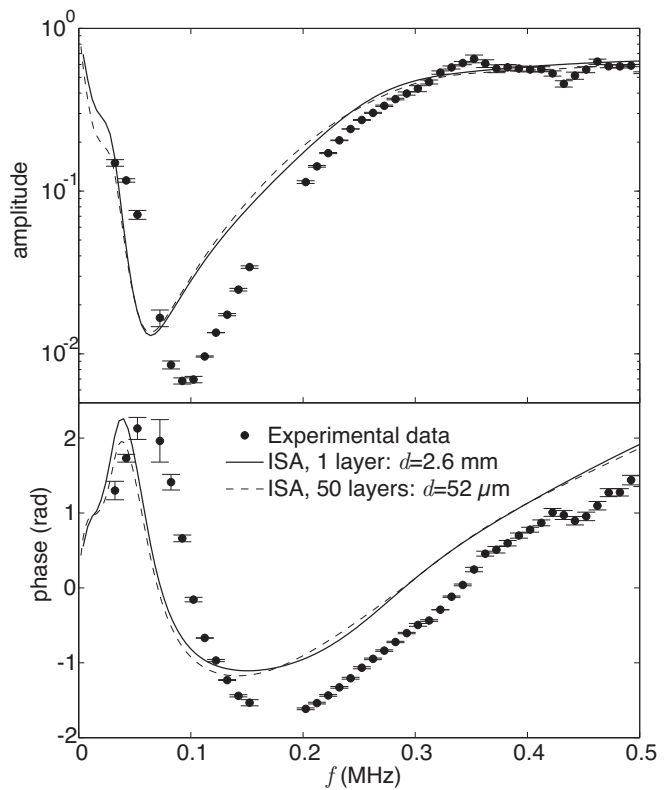}
\caption{\textit{(Color online) Amplitude (top) and phase (bottom) of the acoustic transmission through the sample. Because the data are the same as in figure~\ref{fig-fitPVAt}, the same discrepancy between Foldy's prediction for a homogenous sample (solid lines) and the experimental data (circles) is obtained. The prediction for a layered medium (dashed lines) does not bring better agreement.}\label{fig-T}}
\end{center}
\end{figure}

We {investigated }
the influence of this heterogeneity of the sample on the acoustic propagation. Each layer i ($1<i<50$) of thickness $d=52\,\mum$ was considered as an effective medium following Foldy's model with a concentration $n^\prime/d$ and a bubble size distribution as measured by the x-ray tomography. Following Brekhovskikh~\cite{Bre1960}, we computed the transmission through this multi-layer system, and compared it to the experimental transmission. Figure~\ref{fig-T} shows the result: accounting for the inhomogeneity of the bubble concentration with the $50$ layer model does not change the prediction much, implying that the heterogeneity of the bubble concentration plays a negligible role in acoustic wave propagation through the medium.

\subsection{Positional correlations}\label{sec-correlations}
Another approximation in Foldy's model {is the assumption that }
the positions of the scatterers
{ are }uncorrelated. Recent experiments in a $2$D system of {steel }
rods in water have shown that Keller's approach results in a very good correction {to the model} when correlations needed to be included~\cite{Derode2006}. In Keller's approach, the effective wave vector is related to the correlation function by:
\begin{eqnarray}
k^2 = k_0^2 &+& 4\pi n f_\text{s}\nonumber \\
& -& (4\pi n f_\text{s})^2 \int (1-g(r))\frac{\sin kr}{k} \text{e}^{\ii k_0r} \ud r\label{eq-keller}
\end{eqnarray}
where the correlations are taken into account by a radial function $g(r)$ such that the local concentration {of }
scatterers at a distance $r$ from a scatterer is $n(r)=n_0 g(r)$, where $n_0$ is the average concentration. {For }
point-like scatterers, $g=1$ and Foldy's approximation is recovered. For hard spheres, $g=0$ for $r<2R$ and $1$ at larger distances. In general, $g$ can take more complicated forms, depending on the correlation mechanisms involved in the medium.
\begin{figure}[ht]
\begin{center}
\includegraphics[width = \linewidth]{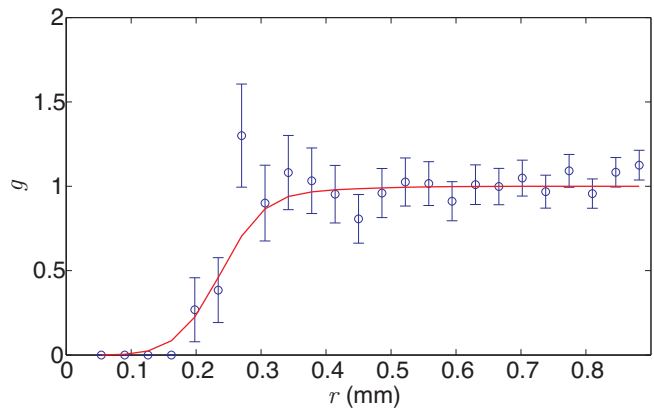}
\caption{\textit{(Color online) Function $g$ as measured from the x-ray tomography data (circles). Error-bars correspond to one standard deviation. The solid line corresponds to the hard-sphere approximation for the bubble size distribution measured in the sample.}\label{fig-g}}
\end{center}
\end{figure}

From the x-ray tomography data, we were able to estimate the function $g$ for our sample. The general procedure for measuring such a function {involves the following steps. First, one considers }
a sphere of radius $r_N$ around one bubble and count{s} 
the number of bubbles in each spherical layer between $r_i$ and $r_{i+1}$. Then, averaging over different central bubbles gives a statistical estimate of $g(r)$. However, a limit of this technique is that the bubbles chosen as centres of the spheres cannot be at less than $r_N$ from an edge of the sample. Thus, if one wants to investigate long-range correlations by taking high value{s} of $r_N$, the statistics {are }
poor because the number of bubbles that are far enough from the borders is small. For our thin sample, this limitation was severe. To circumvent this problem, we used $1/8$ of each sphere, with an orientation chosen so that its volume did not cross any boundaries of the sample. Figure~\ref{fig-g} shows the result of {these measurements up to }
a maximum distance of $0.9\,$mm. It appears that the $g$ function is very similar to {the predictions of the }
hard-sphere approximation 
(solid line). Note that the oscillations in the $g$ function that are predicted by the {frequently used }Perkus-Yevick approximation~\cite{Derode2006} are not expected to be significant here because the concentration of scatterers is low. Numerical simulations with random positions of scatterers in a box confirm that a simple hard-spheres law is a good approximation for a $2\,\%$ volume fraction.

\begin{figure}[ht]
\begin{center}
\includegraphics[width = \linewidth]{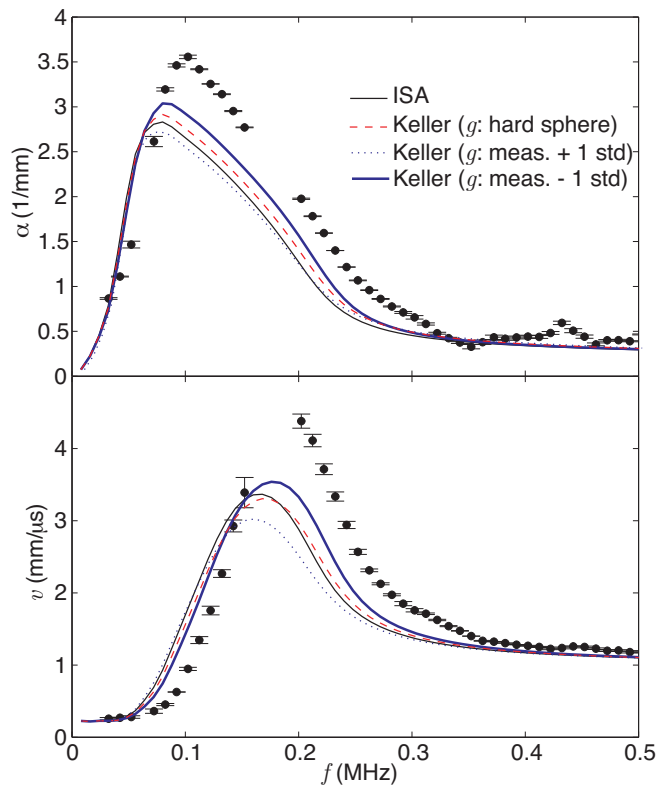}
\caption{\textit{{(Color online) Investigation }
of the effect of correlations. The experimental attenuation and velocity (circles) and the ISA prediction (solid thin lines) are shown again, as in figure~\ref{fig-PVAt}. Keller's approach (Eq.~(\ref{eq-keller})) is plotted for the hard-sphere correlation (dashed lines), for the high estimation of $g$ measured from the x-ray data (dotted lines), and for the low estimation of $g$ (thick solid lines).}\label{fig-corr}}
\end{center}
\end{figure}

The predictions of Keller's model for the attenuation and velocity in the sample are shown in Fig.~\ref{fig-corr} for three different $g$ functions. When the hard sphere approximation is used (\emph{i.e.}, the solid curve of Fig.~\ref{fig-g} is taken), one obtains the dashed lines, which only give a slightly better agreement than the ISA. From the measured correlations, we also consider
the high limit (\emph{i.e.}, the experimental points plus one standard deviation) and the low limit (\emph{i.e.}, the experimental points minus one standard deviation) for the $g$ function. They give different corrections: while the former gives a worse agreement, the latter improves it. One can understand {this difference qualitatively by re-writing} equation~(\ref{eq-keller}) as
\begin{eqnarray}
k^2 = k_0^2\left[1 + \frac{4\pi n f_\text{s}}{k_0^2} \left(1 -4\pi n f_\text{s}\int (1-g)  r \ud r \right) \right], \label{eq-keller2}
\end{eqnarray}
where the reasonable approximations $kr\ll 1$ and $k_0 r\ll1$ have been made. It is important to note that, in the frequency range over which the discrepancy is high, {the }bubbles are responding in phase opposition to the pressure field, \emph{i.e.} $\text{Re}(f_\text{s})<0$. So, according to equation~(\ref{eq-keller2}), the correlations increase the effect of the bubbles on $k$ if $g<1$, whereas they decrease it if $g>1$. Note that a simple criterion can be deduced from Eq.~(\ref{eq-keller2}) to estimate the importance of the corrections due to correlation. If $g$ is roughly described as a step function going from $0$ to $1$ at distance $r_\text{c}$, the term with {the }
integral in Eq.~(\ref{eq-keller2}), which we will denote as $\beta$, reduces to
\begin{eqnarray}
\beta = \frac{3}{2}\Phi \frac{f_\text{s}}{R}\left(\frac{r_\text{c}}{R}\right)^2. \label{eq-corcor}
\end{eqnarray}
Thus, the magnitude of the correction depends on the ratio of $r_\text{c}$ to the radius of the bubbles $R$. Long-range correlations are thus expected to have a stronger effect.

\begin{figure}[ht]
\begin{center}
\includegraphics[width = \linewidth]{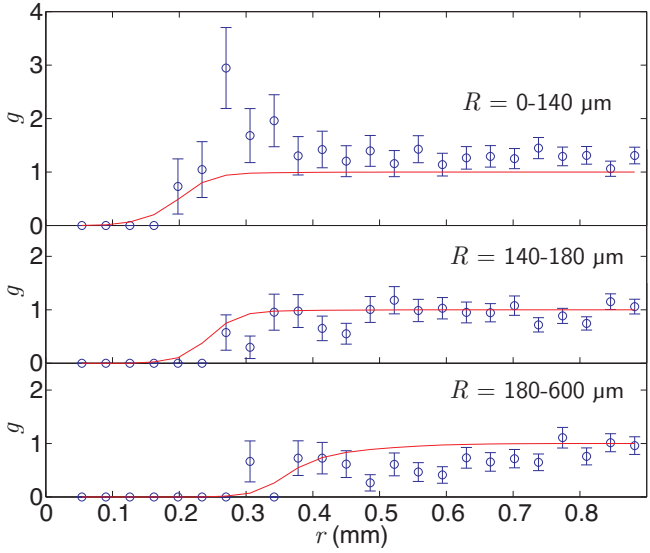}
\caption{\textit{(Color online) Function $g(r,R)$ as measured from the x-ray tomography data (circles) on three populations of bubble radii. Error-bars correspond to one standard deviation. The solid line corresponds to the hard-sphere approximation for the bubble size distribution measured in the sample.}\label{fig-g3}}
\end{center}
\end{figure}

It appears that Keller's approach, with {a }reasonable estimat{e }
of the $g$ function, {introduces }
a correction that goes in the right direction for better agreement. But the agreement is still unsatisfactory. {Moreover}
, Keller's expression applies to a monodisperse medium, whereas our bubbly PDMS is a polydisperse sample. So far, we have assumed that the polydispersity could be taken into account by changing $n f_\text{s}$ into $\int n(R)\ud R f_\text{s}(R)$ in the equations. However, as we show in appendix~\ref{annexe}, a self-consistent approach provides a more general expression (see equation~(\ref{CorreMoiPoly})), in which the correlations 
depend on the size of the scatterer: $g(r)$ becomes $g(r,R)$.
To examine this {polydispersity }effect
, we considered $3$ different populations of ``central'' bubbles: $0$-$140\,\mum$, $140$-$180\,\mum$, and $180$-$600\,\mum$. For each population, we were able to average over approximately $1000$ bubbles, ensuring good statistics.
Figure~\ref{fig-g3} shows the $g$ functions obtained for these $3$ populations. For the small bubbles, there is a clear maximum of the $g$ function. This is an indication of clusters. A close inspection of the picture in figure~\ref{fig_photo} indeed gives the impression that the small bubbles are clustered. For the average size bubbles, the $g$ function is similar to the hard-sphere approximation. For the big bubbles, we find evidence of depletion: there is a lower probability of finding a bubble close to a big one. Again, note that the photo in Fig.~\ref{fig_photo} shows isolated big bubbles.

\begin{figure}[ht]
\begin{center}
\includegraphics[width = \linewidth]{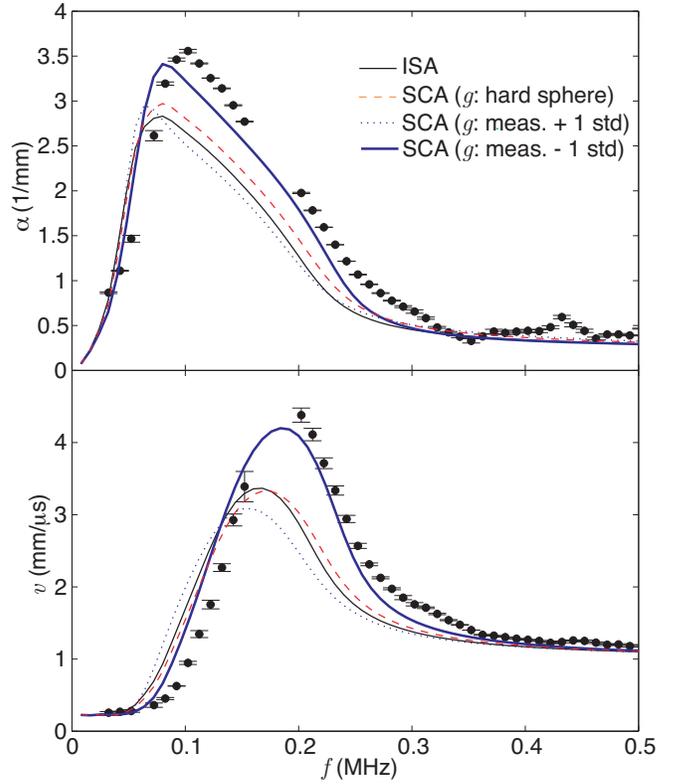}
\caption{\textit{
(Color online) The effect of polydisperse correlations as predicted by the self-consistent approach (SCA).
Three $g(R,r)$ functions are considered (see figure~\ref{fig-g3}): hard-sphere correlation (dashed lines), high and low estimates
of $g$ measured from the x-ray data (dotted  and thick lines, respectively).}\label{fig-corr2}}
\end{center}
\end{figure}

Figure~\ref{fig-corr2} shows the predictions of our 
self-consistent approach (SCA) when $3$ different possibilities are considered for $g(r,R)$: hard sphere approximation (solid lines in Fig.~\ref{fig-g3}), and the measured values of $g$ including the low and the high limits (bottom and top of the error bars in Fig.~\ref{fig-g3}).
It appears that the effects observed in Fig.~\ref{fig-corr} are amplified when the polydisperse correlation is taken into account. Interestingly, even for the hard sphere approximation, slightly better agreement is found (see table~\ref{Tab3}). This can be explained by the following argument: large bubbles scatter more than small bubbles (see equation~(\ref{eqf})), and their average correlation distance $r_\text{c}$ is also larger.
\begin{table}[htdp]
\caption{\textit{A simple criterion for estimating how well the predictions of the different models agree with the experimental data: the relative mean squared differences between theoretical and experimental effective wavenumbers (average of the imaginary and real parts) over the $30$-$300\,$kHz frequency range. Different $g$ functions are considered: from the hard-sphere approximation (hs), and from the measured correlation (see Fig.~\ref{fig-g3}) with the lower (xp-) middle (xp) and upper (xp+) estimates.}}
\begin{center}
\begin{tabular}{|c | c| c c c c| c c c c|}
\hline
  Model & ISA  & \multicolumn{4}{|c|}{Keller} & \multicolumn{4}{|c|}{SCA}\\
  \hline
  \textbf{$g$} & / & hs & xp - & xp & xp + & hs & xp - & xp & xp + \\
 \hline
\textbf{($\%$)}& $14.8$ & $11.3$ & $7.3$ & $12.4$ & $17.7$ & $10.1$ & $4.3$ & $10.8$ & $19.6$ \\
\hline
\end{tabular}
\end{center}
\label{Tab3}
\end{table}%

When the range of measured values of $g$ is considered, the agreement is better or worse, depending on whether the high or the low limit of the possible $g$ function is considered. It follows that, given the precision of our correlation measurements, we cannot conclude whether or not the polydisperse correlations are sufficient to fully explain the discrepancy. Nonetheless, we find that plausible values for $g$ are able to give good agreement. A more precise determination of the correlations would be necessary for a definitive conclusion to be drawn.

\section{Conclusions}
We have investigated wave propagation through a complex medium in which low-frequency scattering resonances lead to strong dispersion, with large peaks in the velocity and attenuation. By choosing to work with a relatively simple acoustic system -- bubbles in the elastic medium PDMS -- we have shown how these signatures of strong resonant scattering are influenced by the coupling between the scatterers.  This leads to the breakdown of the  independent scatterer approximation (ISA), which is commonly used to interpret experimental data.  A crucial step in investigating the failure of the ISA in this system has been our ability to characterize the physical properties of bubbly PDMS very carefully, including independent measurements of the elastic properties of the matrix as well as the concentration and size distribution of the scatterers. Remarkably, we find that this failure of the ISA occurs even though the concentration of scatterers is low, only $2 \%$ in our case.  It is worth noting one important consequence: if the ISA is used to estimate the size distribution of the scatterers, significantly smaller sizes are found than by direct imaging methods.


 In this paper, we have proposed that this discrepancy can be explained by accounting for the role of correlations in the scatterers' positions, which we were able to probe directly using x-ray tomography. Following Keller's approach, we have shown that these correlations brought a non-negligible correction to the ISA, and that the agreement with the experimental data was improved, although not perfect. We have proposed an alternative approach, based on a self-consistent argument, with which we were able to take into account the effect of polydispersity on the correlations: bubbles of different radii are not correlated with their neighbours in the same way. Again, x-ray tomography allowed us to investigate
 this polydisperse correlation, which was found to be different from the monodisperse estimate: small bubbles are more likely to be close to other bubbles, whereas big bubbles are often isolated. We have shown that, within the possible polydisperse correlations that are consistent with our x-ray measurements, the SCA was able to provide a satisfactory explanation for the observed shift to higher frequencies in the acoustic resonances in the sample. 
It should also be noted that our model could be refined to account for more complex situations. 
In particular, when the correlations are long-range in a thin sample, one can expect the boundaries to play a role in the correlation, an effect we did not incorporate. Also, we looked at correlation effects or inhomogeneity effects, but we have not considered the case in which both effects are coupled.

An interesting question is the following: why are 
correlations important in the bubbly medium studied here, whereas they have not been detected before for bubbles in water~\cite{Wilson2005} or bubbles in a yield stress fluid~\cite{Leroy2008}? According to equation~(\ref{eq-corcor}), the magnitude of the 
correction due to the positional correlations is proportional to the volume fraction of scatterers $\Phi$, which was generally lower in previous studies ($5\times10^{-4}$ to $10^{-2}$ at most), so that $\beta$ would be expected to be low. However, it is interesting to note that some previous studies did report deviations between Foldy's prediction and the experimental data that might be interpreted as a manifestation of correlation effects. In the historical data by Silberman~\cite{Sil57}, for instance, the measured attenuation was much larger than Foldy's estimate. Feuillade proposed that the discrepancy could be resolved 
by introducing an arbitrary fitting parameter~\cite{Feu96} that one can interpret 
as a correlation length 
(see the appendix). Wilson \emph{et al.} also observed deviations from Foldy's estimate, as their attenuation peak 
was at a lower frequency than what was expected from the size analysis (see Fig.~$10$ in \cite{Wilson2005}). They attributed this discrepancy to the uncertainty 
in their measurement of the bubble size distribution, but one can imagine that correlation effects were responsible for this shifting of the attenuation peak
. Indeed, large 
depletion zones around bubbles have been observed in bubbly liquids~\cite{Cartellier2001}, which would lead to positional correlations. In our sample, the exact mechanism leading to the positional correlations has not been identified unambiguously. Ostwald ripening might induce the disappearance 
of small bubbles close to bigger bubbles, hence leading to the isolated big bubbles we observed. Another possible scenario could involve the rotation of the cell when the sample was prepared. If one considers that only the buoyancy and the Stokes drag force are relevant (inertial forces were negligible in our set-up), a bubble is expected to move along a circle whose radius is inversely proportional to the rotation frequency, and proportional to the square of the bubble radius. The movement of the bubbles may induce hydrodynamic interactions leading to positional correlations. This hypothesis is supported by the observation of strong clustering when the cell was rotated at a lower speed ($0.7\,$rpm), \textit{i.e.}, when the radii of the circles followed by the bubbles were larger.

\begin{acknowledgments}
Support from the Natural Sciences and Engineering Research Council
of Canada is gratefully acknowledged.
The authors wish to thank A. Derode, M. Devaud, T. Hocquet, and J.-C. Bacri for numerous discussions, as well as
R. Jayaraman for his help in {enabling the }
x-ray experiments {to be carried out}.
\end{acknowledgments}

\appendix
\section{Self-consistent approach for multiple scattering of waves}\label{annexe}
{We propose a self-consistent approach for calculating the dispersion relation of a random multiply scattering medium, taking into account the positional correlations of the scatterers.} Let us consider an infinite medium with $n$ inclusions per unit volume. We limit ourselves to low frequencies such that the wavelength is large compared to the size of the {scattering} inclusions, which means we can consider that the waves scattered by the inclusions are spherical: when scatterer $i$ is excited {by a wave} with 
amplitude $p_i$, it generates a wave  $p_i f_\text{s}^i \text{exp}(\ii k_0 r)/r$ at distance $r$, where $f_\text{s}^i$ is its scattering function and $k_0$ the wave number in the pure medium. The total field experienced by inclusion $i$, due to the other scatterers, is given by the self-consistent relation
\begin{eqnarray}
p_i =  \sum_{j\neq i}  p_j f_\text{s}^j \frac{\text{e}^{\ii k_0 r_{ij}}}{r_{ij}}, \label{eq1}
\end{eqnarray}
where $r_{ij}$ is the distance between scatterers $i$ and $j$. In the case of a finite number of scatterers, Eq.~(\ref{eq1}) can be solved and all the $p_i$ exactly computed. For an infinite number of inclusions, however, taking into account the full multiple scattering is a difficult task and the $p_i$ are not easily accessible. In particular, the $1/r$ range of the interaction means that no {pair} 
($i,j$) can be 
neglected {\textit{a priori}} in Eq.~(\ref{eq1}).

\begin{figure}[ht]
\begin{center}
\includegraphics[width = .6\linewidth]{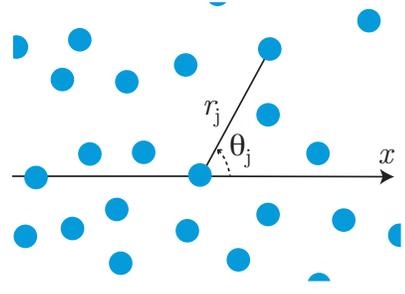}
\caption{\textit{(Color online) An infinite $3D$ medium with spherical scatterers is considered. The position of scatterer $0$ is arbitrarily chosen as the centre of the coordinates and we look for modes of the system consisting of plane waves propagating in the $x$ direction. The dispersion relation is obtained by calculating the total field experienced by scatterer $0$ due to the other scatterers.}\label{fig-notation}}
\end{center}
\end{figure}

Let us further simplify the problem by assuming that all the scatterers are identical ($f_\text{s}^i= f_\text{s}$) and look for modes that take the form of plane waves propagating in the $x$ direction: $p_i=P \text{exp}(\ii k x_i)$, where $k$ is the effective wave number we want to determine, and $P$ the amplitude of the mode. Then, Eq.~(\ref{eq1}) reduces, for scatterer $i=0$ arbitrarily chosen as the central one {located at $r=0$}, to
\begin{eqnarray}
1 &=&  \sum_{j\neq 0}  \text{e}^{\ii k x_j} f_\text{s} \frac{\text{e}^{\ii k_0 r_{j}}}{r_{j}} \nonumber \\
&=&f_\text{s} \sum_{j\neq 0} \frac{\text{e}^{\ii(kr_j\text{cos}\theta_j + k_0 r_j)}}{r_j}, \label{eq1b}
\end{eqnarray}
where $r_j$ is the distance to scatterer $j$, and $\theta_\text{j}$ the angle between the direction of propagation and the position of scatterer $j$ (see {Fig.}
~\ref{fig-notation}).

 One can approximate the discrete equation by a continuous equation, taking into account the {positional} 
 correlations with the pair correlation function $g(r)$:
\begin{subequations}
\begin{eqnarray}
1&=& f_\text{s} \int  \ud^3 r \times ng(r) \frac{\text{e}^{\ii(kr\text{cos}\theta + k_0 r)}}{r} \label{eqA} \\
&=&\frac{4\pi n f_\text{s}}{k} \int_0^\infty ng(r)\sin(kr) \text{e}^{\ii k_0 r}\ud r.
\end{eqnarray}
\end{subequations}

Invoking a small imaginary part of $k_0$ to preserve the convergence of the integral, one can then calculate the effective wave number:
\begin{eqnarray}
k^2 = k_0^2 + \frac{4\pi nf_\text{s}}{1+4\pi nf_\text{s} \int (1-g(r)) \frac{\sin kr}{k}\text{e}^{\ii k_0r} \ud r}.\label{CorreMoi}
\end{eqnarray}
Note that equation~(\ref{CorreMoi}) reduces to Foldy's equation if the denominator is approximated by $1$, and to Keller's equation if it is expanded to second order in $nf_\text{s}$. Interestingly, it has some similarities with the equation proposed by Feuillade, which was found to give excellent agreement with experimental data for bubbly liquids (see equation ($30$) in \cite{Feuillade1997}). Feuillade's model has been subject to criticisms~\cite{Ye1997, Hen99}, in particular because, for unclear reasons,  it assumes that the interactions between the scatterers are confined within a finite range. It might be that this finite range is related to correlations. Note that the approach we develop here is similar to Feuillade's. The differences are that ($1$) we abandon the assumption that all the scatterers are in phase, which is true only locally, and ($2$) we include the positional correlations.

For a polydisperse assembly of scatterers, the same self-consistent scheme can be used. In this case, the total field $p_i$ 
depends \textit{a priori} on the radius of the bubble, $p_i=P(R_i) \text{exp}(\ii k x_i)$, meaning that equation~(\ref{eqA}) must be generalized into
\begin{eqnarray}
P(R) &=&f_\text{s}(R) \int n(R^\prime) \ud R^\prime P(R^\prime)\times \nonumber \\
&&  \int g(r,R,R^\prime) \frac{\text{e}^{\ii(kr\text{cos}\theta + k_0 r)}}{r}\ud^3 r \label{eqAb},
\end{eqnarray}
where $g(r,R,R^\prime)$ gives the probability of finding a bubble with radius $R^\prime$ at a distance $r$ from a central bubble with radius $R$. In the general case, equation~(\ref{eqAb}) cannot be directly simplified into a dispersion relation. However, if we assume that the correlation does not depend on $R^\prime$ ($g(r,R,R^\prime)=g(r,R)$), a further integration over $R$ yields
\begin{eqnarray}
1 &=&\int n(R)f_\text{s}(R) \ud R\int g(r,R) \frac{\text{e}^{\ii(kr\text{cos}\theta + k_0 r)}}{r}\ud^3 r \label{eqAc},
\end{eqnarray}
 from which we obtain the dispersion relation
 \begin{eqnarray}
 k^2 = k_0^2 +  \frac{\int 4\pi n(R)f_\text{s}(R) \ud R }{1+A},\label{CorreMoiPoly}
\end{eqnarray}
with
 \begin{eqnarray}
A= \int  4\pi n(R)f_\text{s}(R) \ud R  \int (1-g(r,R)) \frac{\sin kr}{k}\text{e}^{\ii k_0r} \ud r. \nonumber
\end{eqnarray}
Note that, as in the ISA, 
polydispersity modifies the equation by changing $nf_\text{s}$ into $\int n(R) \ud R f_\text{s}(R)$. But polydispersity also brings another non-trivial modification: the correlation can depend on the radius of the central bubble considered, which makes different correlations possible for different bubble sizes, as in our bubbly sample.

\end{document}